\documentclass[aps,prl,twocolumn,amsmath,amssymb,nofootinbib,superscriptaddress,floatfix,reprint,longbibliography]{revtex4-1}
\usepackage[dvips]{graphicx}
\usepackage{latexsym}
\usepackage{amsmath}
\usepackage{amsfonts}
\usepackage{amssymb}
\usepackage{color}
\usepackage{txfonts}
\usepackage{float}
\usepackage{url}
\usepackage[colorlinks=true, urlcolor=blue, linkcolor=blue, citecolor=blue]{hyperref}
\usepackage{ulem}
\usepackage{tikz-feynman}
\usepackage{tikz,tikz-feynhand} 
\usepackage{graphicx}
\usepackage{extpfeil}
\usepackage{subfigure}
\usepackage{physics}
\usepackage{titlesec}
\titleformat{\section}{\normalfont\normalsize\bfseries\centering}{\thesection.}{1em}{}
\begin{document}
	\newcommand{\fig}[2]{\includegraphics[width=#1]{#2}}
	\newcommand{\la}{{\langle}}
	\newcommand{\ra}{{\rangle}}
	\newcommand{\dg}{{\dagger}}
	\newcommand{\upa}{{\uparrow}}
	\newcommand{\dna}{{\downarrow}}
	\newcommand{\ab}{{\alpha\beta}}
	\newcommand{\ias}{{i\alpha\sigma}}
	\newcommand{\ibs}{{i\beta\sigma}}
	\newcommand{\hH}{\hat{H}}
	\newcommand{\hn}{\hat{n}}
	\newcommand{\hc}{{\hat{\chi}}}
	\newcommand{\hU}{{\hat{U}}}
	\newcommand{\hV}{{\hat{V}}}
	\newcommand{\br}{{\bf r}}
	\newcommand{\bR}{{\bf R}}
	\newcommand{\bA}{{\bf A}}
	\newcommand{\bk}{{{\bf k}}}
	\newcommand{\bq}{{{\bf q}}}
	\newcommand{\ri}{{\mathrm{i}}}
	\def\gsim{~\rlap{$>$}{\lower 1.0ex\hbox{$\sim$}}}
	\setlength{\unitlength}{1mm}
	\newcommand{\pprl}{Phys. Rev. Lett. \ }
	\newcommand{\pprb}{Phys. Rev. {B}}

\title {The coherence peak of unconventional superconductors in the charge channel}
\author{Pengfei Li}
\affiliation{Beijing National Laboratory for Condensed Matter Physics and Institute of Physics,
	Chinese Academy of Sciences, Beijing 100190, China}
\affiliation{School of Physical Sciences, University of Chinese Academy of Sciences, Beijing 100190, China}

\author{Zheng Li}
\email{lizheng@iphy.ac.cn}
\affiliation{Beijing National Laboratory for Condensed Matter Physics and Institute of Physics,
	Chinese Academy of Sciences, Beijing 100190, China}
\affiliation{School of Physical Sciences, University of Chinese Academy of Sciences, Beijing 100190, China}

\author{Kun Jiang}
\email{jiangkun@iphy.ac.cn}
\affiliation{Beijing National Laboratory for Condensed Matter Physics and Institute of Physics,
	Chinese Academy of Sciences, Beijing 100190, China}
\affiliation{School of Physical Sciences, University of Chinese Academy of Sciences, Beijing 100190, China}

\date{\today}

\begin{abstract}
In this work, we carry out a systematic investigation of the coherence peak in unconventional superconductors as they transition into the superconducting phase at $T_c$. Using $d$-wave cuprates as an example, we reveal the presence of a coherence peak below $T_c$ in the charge channel. The nuclear quadrupole relaxation rate is shown to be an effective method for detecting this unconventional coherence peak, with the superconducting coherence factor playing a pivotal role in its emergence. Additionally, we explore the influence of correlation effects, which further enhance this phenomenon. Extending our analysis, we demonstrate the existence of a similar coherence peak in ultrasonic attenuation and iron-based superconductors. Our findings offer a fresh perspective on probing superconducting gap symmetry in unconventional superconductors.
\end{abstract}

\maketitle
\begin{figure}[t]
	\begin{center}
		\fig{3.4in}{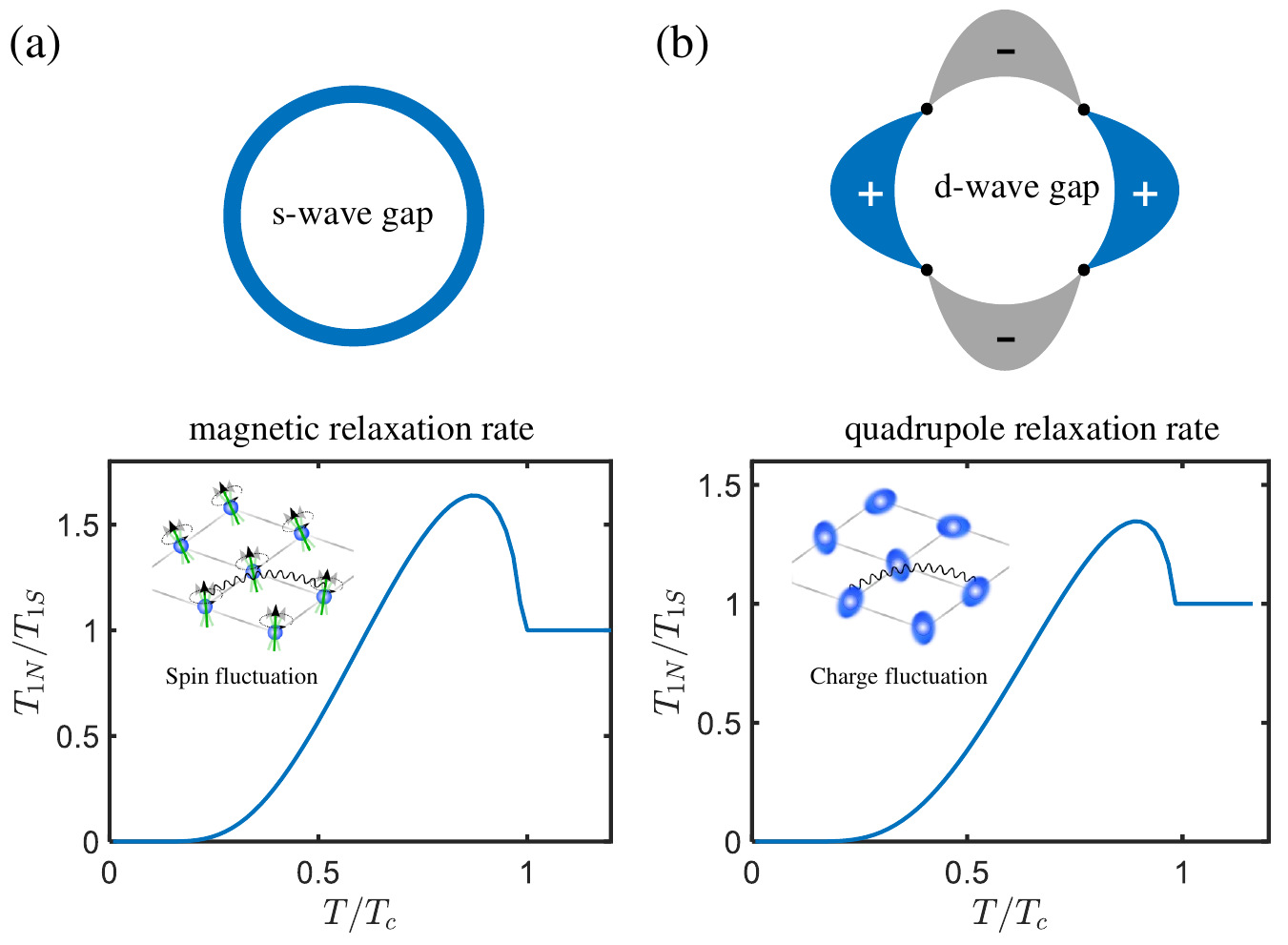}
		\caption{(a) Top: the gap function of $s$-wave superconductors. Bottom: the spin-lattice relaxation rate of conventional superconductors. Owing to the diverged quasiparticle density of states, the spin fluctuation-related relaxation rate acquires the Hebel-Slichter coherence peak crossing $T_c$.
        At lower temperature, few quasiparticles are excited and the relaxation rate goes down to zero.
        (b) Top: the gap function of $d$-wave superconductors. Bottom: the quadrupole relaxation rate of cuprates superconductors. This charge fluctuation-related relaxation rate also acquires a new coherence peak crossing $T_c$.
			\label{fig1}}
	\end{center}
		\vskip-0.5cm
\end{figure}

Owing to electron-electron correlation, the emergency of high-temperature (high-$T_c$) superconductors and other related unconventional superconductors greatly challenges our understanding based on the conventional Bardeen–Cooper–Schrieffer (BCS) theory \cite{keimer_review,doping_mott,schrieffer2007handbook}.
Although solving this complex correlation problem seems to be a formidable task, uncovering their unconventional properties is one achievable work and an inevitable step toward the final destination \cite{keimer_review}.
Retracing the development of conventional superconductors and BCS theory \cite{bcs_theory,schrieffer}, coherence peaks always served as barometers for understanding their physical properties. For example, the coherence peak from the scanning tunneling microscope (STM) determines the diverged density of states (DOS) at the superconducting gap \cite{bcs_theory,schrieffer}; the Hebel-Slichter coherence peak \cite{HS_PhysRev.113.1504,schrieffer}, observed in the nuclear magnetic resonance (NMR) spin-lattice relaxation rate 1/$T_1$ just below the transition temperature $T_c$, as illustrated in Fig.\ref{fig1}(a). This spin-fluctuation-related coherence peak is one important hallmark for the $s$-wave spin-singlet pairing supporting the BCS theory \cite{HS_PhysRev.113.1504,schrieffer}. 

For cuprates and iron-based superconductors, coherence peaks have already played important roles in extracting their electronic structures \cite{keimer_review,scalapino_RevModPhys.84.1383,schrieffer2007handbook,pengchengdai_review,doping_mott}. For example,
neutron scattering experiments observed the spin resonance peaks at $Q=(\pi,\pi)$ in cuprate superconductors \cite{Spin_ROSSATMIGNOD199186,Spin_PhysRevLett.75.316,Spin_PhysRevLett.70.3490}, which revealed the novel spin dynamics for high-T$_c$ supercondcutors.
On the other hand, NMR failed to locate any coherence peaks entering superconducting states in cuprates, which is beyond the conventional understanding of BCS theory \cite{A_Rigamonti_1998,nmr_ybco_PhysRevLett.59.1860,nmr_ybco_PhysRevB.36.5727,fukuyama_nmr,pines_PhysRevB.41.6297,scalapino_nmr,Tachiki_PhysRevB.39.2279,Scalapino_PhysRevLett.68.706,Scalapino_PhysRevLett.67.2898}. Recently, we also successfully extracted a new coherence peak from the quadrupole relaxation rate $1/T_1^{\rm quad}$ \cite{NQR_PhysRevX.14.041072}, as illustrated in Fig.\ref{fig1}(b). 
Importantly, this peak is absent in the BCS superconductors pointing to unconventional charge fluctuations crossing $T_c$ in $d$-wave superconductors. This coherence peak provides a new barometer for unconventional superconducting pairing \cite{NQR_PhysRevX.14.041072}.
Motivated by this observation, we perform a systematic study on the coherence peak in unconventional superconductors through the charge fluctuation channel in this work. 

Traditionally, NMR focuses on the interaction between the nuclear spin and electron spin \cite{slichter2013principles}. 
During the process non-equilibrium nuclear spin state acquiring thermodynamic equilibrium with its surrounding electrons, the nuclear spin relaxation time $T_1$ provides direct information on the spin fluctuation of surrounding electrons \cite{slichter2013principles}. 
On the other hand, the nonspherical nuclear spin with the spin $I>1/2$ also hosts nuclear quadrupole moment, which couples the electron charge densities through the electric field gradient (EFG) at nuclear sites \cite{slichter2013principles}. 
Therefore, the spin-lattice relaxations contain both magnetic relaxation by the electron spin fluctuation and the quadrupole relaxation by the electron charge fluctuation.
This quadrupole relaxation is a highly accurate measurement of charge degree of freedom \cite{slichter2013principles,nqr_mitchell,NQR_PhysRev.104.271}. 

Phenomenologically, the coupling between the EFG and the nuclear quadrupole moment can be written as 
\begin{eqnarray}
    H_{Q}=\sum_{q} F(q) \hat{\rho}(q) \{3\hat{I}_z^2-\hat{I}^2+\frac{\eta}{2}(\hat{I}_{+}^2+\hat{I}_{-}^2) \}(-q)
\end{eqnarray}
where $F(q)$ is the structure factor. $\hat{\rho}(q)$ is the electronic density operator and $\hat{\mathbf{I}}$ is the nuclear spin operator. $\eta=(V_{xx}-V_{yy})/V_{zz}$ is the EFG asymmetry parameter defining the difference between the EFG tensors $V_{xx}$, $V_{yy}$ and $V_{zz}$ \cite{slichter2013principles}.
Following the same procedure of magnetic relaxation rate in NMR \cite{slichter2013principles,xiang2022d,pines_PhysRevB.41.6297}, the quadrupole relaxation rate $1/T_1$ \cite{nqr_mitchell,NQR_PhysRev.104.271,nqr_theory} can be written as
\begin{equation}
    \frac{1}{T_1} \propto T\sum_q F^2(q)\lim_{\omega\to 0} \frac{\Im \chi(q,\omega)}{\omega}.
\end{equation}
where $\chi$ is the electron charge susceptibility.
Normally, the spatial dependence of $F(q)$ is weak as in cuprates \cite{YBCO_PhysRevB.42.2051} and we can take $F(q)$ as a constant in most cases.

\begin{figure}[t]
	\begin{center}
		\fig{3.4in}{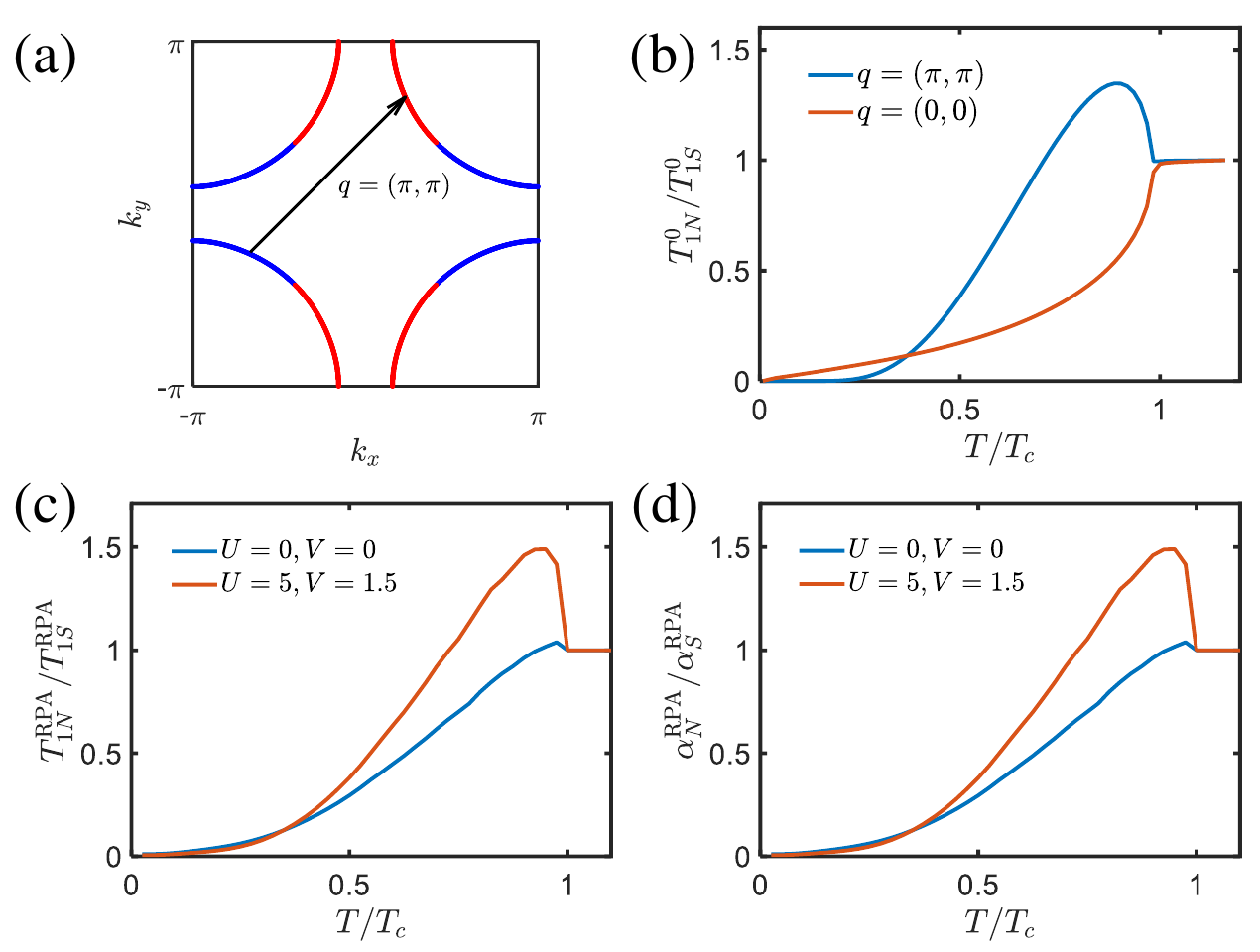}
		\caption{(a) The Fermi surface of cuprates with $d$-wave pairing symmetry. The colors represent the sign of order parameter on the FS with red indicating positive and blue indicating negative. 
        (b) The bare quadrupole relaxation rate (non-interacting) of cuprates at $q=(\pi,\pi)$ and $q=(0,0)$. $q=(0,0)$ shows the same feature as a $s$-wave superconductors while the $q=(\pi,\pi)$ shows a coherence peak owing to the coherence factor.  
        (c) The total quadrupole relaxation rate calculated based on charge susceptibility for the bare noninteracting case (blue line) and for the RPA renormalized case (orange line). The coherence peak is enhanced at the orange line below $T_c$.
     (d) The ultrasonic attenuation rates are another way of probing the charge channel, which shows features similar to the quadrupole relaxation rate in (c).
			\label{fig2}}
	\end{center}
		\vskip-0.5cm
\end{figure}

Without losing the generality, we first take the one-band model for cuprates as one example and discuss other cases later. 
The mean field Hamiltonian for this one-band model can be written as
\begin{eqnarray}
H_0 = \sum_{k\sigma}\xi_k c^\dagger_{k\sigma}c_{k\sigma} + \Delta_k c^\dagger_{k\uparrow}c^\dagger_{-k\downarrow} + \mathrm{H.c.}
\end{eqnarray}
where $\xi_k=-2t(\cos{k_x}+\cos{k_y})+4t'\cos{k_x}\cos{k_y}-\mu$ is the normal state dispersion. We set $t=1$, $t'=0.35$ and $\mu=-1$ to capture the feature of Fermi surface (FS) in hole-doping cuprates. 
It is well-known that cuprates pairing symmetry belongs to $B_{1g}$ \cite{Tsuei_RevModPhys.72.969}. Hence, the pairing gap function takes $\Delta_k=2\Delta_0(\cos{k_x}-\cos{k_y})$ with $d$-wave form and gap amplitude $\Delta_0$. The $d$-wave gap function can be obtained by solving the gap equation self-consistently with the Fermi surface shown in Fig.\ref{fig2}(a) and the phenomenological nearest neighbor attractive interaction. We will use $\Delta_k$ as a general pairing function for further discussions. The bare charge susceptibility $\chi_0(q,\mathrm{i}\Omega)$ can be calculated through the Green's functions as \cite{xiang2022d} $\chi_0(q,\mathrm{i}\Omega) = \frac{1}{\beta}\sum_{k,\mathrm{i}\omega_n} \Tr \left[\mathcal{G}(k,\mathrm{i}\omega_n)\tau_3\mathcal{G}(k+q,\mathrm{i}\omega_n+\mathrm{i}\Omega_n)\tau_3\right] \label{bare} $,
where $\mathcal{G}(k,\mathrm{i}\omega)$ is the Green's function under Nambu basis $\psi_k=\left[c_{k\uparrow},c^\dagger_{-k,\downarrow}\right]^T$ and $\tau_3$ is the Pauli matrix denoting density operator under Nambu basis. 
To link with experimental probes, the most important quantity is the imaginary susceptibility $\chi_0''(q,\omega)$, which can be written as
\begin{eqnarray}
&&\lim_{\omega\to 0} \frac{\chi_0''(q,\omega)}{\omega}
= \sum_k\int \mathrm{d}\omega f'(\omega)\Tr \left[G^R(k,\omega)\tau_3 G^R(k+q,\omega)\tau_3\right] \nonumber\\
&&= \sum_k\left(1+\frac{\xi_k\xi_{k+q}-\Delta_k \Delta_{k+q}}{E_k E_{k+q}}\right) f'(E_k) \delta(E_k-E_{k+q}) \label{chi0} 
\end{eqnarray}
where $f(\omega)$ is the Fermi distribution function and $E_k=\sqrt{\xi_k^2+\Delta_k^2}$ is the Bogoliubov quasiparticle energy. 
Importantly, the above formula contains one \textit{coherence factor} $g(k,q)=\left(1+\frac{\xi_k\xi_{k+q}-\Delta_k \Delta_{k+q}}{E_k E_{k+q}}\right)$ coming from $(u_ku_{k'}-v_kv_{k'})^2$, where $u_k$,$v_k$ are the mean field Hamiltonian Bogoliubov coefficients \cite{schrieffer}.


One should also notice that Eq.\ref{chi0} is dominated by the Fermi surface contributions with $\xi_k \rightarrow 0$. Hence, for $s$-wave superconductor with 
$\Delta_k=\Delta_0$, $g(k,q)$ around Fermi surface reduces to $1-\frac{\Delta_0^2}{E^2}$.
This $g(k,q)$ almost vanishes for $s$-wave superconductors, which will greatly dismiss the charge fluctuation crossing superconducting transition.
For $s$-wave superconductors, the $1/T_1$ can be expressed as
\begin{equation}
    \frac{1}{T_1} \propto T \int\mathrm{d}E \left(1-\frac{\Delta_0^2}{E^2}\right) N_s^2(E)f'(E)  
    \label{s-wave}
\end{equation}
where $N_s(E)=N_0E/\sqrt{E^2-\Delta_0^2}$ is the density of state (DOS) and $N_0$ is the DOS at $E_F$. 
From Eq.\ref{s-wave}, we can see the DOS divergent factor $1/\sqrt{E^2-\Delta_0^2}$ is compensated by the $E^2-\Delta_0^2$ from the coherence factor $g(k,q)$. Hence, the quadrupole relaxation rate decreases quickly when entering the $s$-wave superconducting state instead of the Hebel-Slichter coherence peak in the magnetic relaxation rate \cite{NQR_PhysRev.104.271,NQR_Ga_PhysRev.120.762,nqr_PSJ.31.1281}.

Moving to the $d$-wave superconductors, the situation becomes more complicated. The gap function $\Delta_k\propto \cos k_x-\cos k_y$ changes sign with $\pi/2$ rotation. 
Hence, the $g(k,q)$ around $q=0$ takes a same feature to $s$-wave superconductors due to $\Delta_{k+q}\Delta_k>0$. $g(k,q)$ around $Q=(\pi,\pi)$ gives rise to an enhancement owing to $\Delta_{k+q}\Delta_k<0$. 
To confirm this, we first take a look at the $1/T_1$ behavior at $q=0$ and $q=Q$ as plotted in Fig.\ref{fig2}(b). The $1/T_1$ has a sudden change at transition temperature $T_c$ as the superconducting gap appears below $T_c$. Here, $T_c$ can be determined by making the eigenvalue of the linearized gap equation to 1.
We can see that the $q=0$ contribution shows a sharp decrease below $T_c$ as for an $s$-wave superconductors. On the contrary, the $q=Q$ contribution acquires a sharp enhancement below $T_c$ as the HS peak.
The sharp peak originates from the $g(k,q)$ around $Q=(\pi,\pi)$ has a different factor from $s$-wave as discussed above. This coherence factor cannot cancel out the DOS logarithmic divergence in the $d$-wave pairing state \cite{xiang2022d}. Notice that, a similar enhancement occurs in the spin fluctuation leading to the neutron scattering spin resonance peak at $Q=(\pi,\pi)$ \cite{Spin_PhysRevLett.70.3490,scalapino_RevModPhys.84.1383}.

However, this physics is more complicated in strongly correlated systems. For the bare susceptibility $\chi_0$, we indeed find a weak coherence peak appearing below $T_c$, as shown in the Fig.\ref{fig2}(c) blue line. Since the quadrupole relaxation rate is calculated by averaging over $q$, it includes contributions from all $q$ points, resulting in the coherent peak being smeared out by the $q\rightarrow0$ decreasing as mentioned above. Hence, to further enhance this peak matching the experimental observations beyond the bare charge susceptibility, 
we should also take the electron correlation effect into account for unconventional superconductors, especially in cuprate high-$T_c$ problem. It is widely believed that the cuprates superconductors are closely related to their Mott parent states \cite{doping_mott}. Hence, the homogeneous charge fluctuation ($q \rightarrow 0$) should be largely quenched.
In order to capture this physics, we take the random phase approximation (RPA) approach to count the corrections to $\chi_0$ from the onsite repulsive interaction $U$ and the nearest-neighbor repulsive interaction $V$. A similar method has been successfully used for magnetic relaxation without Hebel-Slichter coherence peak in cuprates \cite{scalapino_nmr,Scalapino_PhysRevLett.68.706,Scalapino_PhysRevLett.67.2898}.
Using Nambu spinor, the charge vertex $\hat{V}_c$ can be written as
$\hat{V}_c =\frac{1}{2N}\sum_{kk'q} (U/2+V\gamma_q) \psi^\dagger_{k+q}\tau_3\psi_k\psi^\dagger_{k'}\tau_3\psi_{k'+q}$.
where $\gamma_q=2(\cos{k_x}+\cos{k_y})$.
Then the RPA renormalized charge susceptibility can be calculated as
\begin{equation}
    \chi(q,\omega)=\frac{\chi_0(q,\omega)}{1+(U/4+V\gamma_q/2)\chi_0(q,\omega)}.
\end{equation}
In this formula, one can find the $q \rightarrow 0$ fluctuation is largely reduced.
The calculated RPA renormalized quadrupole relaxation rate at $U=5$ and $V=1.5$ is plotted in Fig.\ref{fig2}(c) orange line. The coherence peak is significantly enhanced across $T_c$. 
Hence, we demonstrate that a $d$-wave superconductor could host a coherence peak in its quadrupole relaxation rate.
Similar to the Hebel-Slichter peak in conventional superconductors, the microscopic reason for this quadrupole peak also originates from the DOS logarithmic divergence. The coherence factor and correlation effect further lead to this peak in the charge channel instead of the spin channel.

We want to emphasize that our approach is more or less phenomenological. The main difficulty comes from the unknown pairing mechanism of cuprates and the lack of efficient ways of treating strong correlations.
However, the existence of the coherence peak in the charge channel crossing $T_c$ does not depend on these microscopic details and the way of treating correlation. 
On the other hand, the peak height and the decaying tendency should be sensitive to these details, which calls for further efforts.

There are other ways of detecting pairing properties from the charge channel, like ultrasonic attenuation \cite{schrieffer}.
For conventional superconductors, the longitudinal acoustic attenuation coefficient quickly decreases entering $T_c$ \cite{schrieffer}, which hosts the coherence factor like that in the quadrupole relaxation rate. Experimentally, it has been observed that cuprates superconductors exhibit a peak below $T_c$ \cite{Xu1988, Sun1989}. 
The ultrasonic attenuation rate reflects the scattering rate of phonons by quasiparticles which has the same time reversal symmetry with the quadrupole relaxation rate. Therefore, we can understand these experimental results within the same theoretical framework. The same consideration for $d$-wave superconductor has been explored in Ref.\cite{ultrasonic_PhysRevB.59.7123}. The ultrasonic attenuation rate $\alpha_S$ can be calculated by
\begin{equation}
\begin{aligned}
\alpha_s(T, \mathbf{q})= & A \frac{\Omega}{4 T} \int d \mathbf{k} \frac{1}{\cosh ^2\left(E_{\mathbf{k}} / 2 T\right)} \\
& \times\left(1+\frac{\xi_{\mathbf{k}} \xi_{\mathbf{k}+\mathbf{q}}-\Delta_{\mathbf{k}} \Delta_{\mathbf{k}+\mathbf{q}}}{E_{\mathbf{k}} E_{\mathbf{k}+\mathbf{q}}}\right) \delta\left(E_{\mathbf{k}}-E_{\mathbf{k}+\mathbf{q}}\right),
\end{aligned}
\end{equation}
where $A$ depends on the scattering matrix element and $\Omega$ is the frequency of the ultrasound. Clearly, apart from certain coefficients, this expression is identical to Eq.\ref{chi0}. The ultrasonic attenuation coefficients provided in Ref.\cite{ultrasonic_PhysRevB.59.7123,UA_prl} for different 
$q$ directions are also similar to those in Fig.\ref{fig2}(b). Fig.\ref{fig2}(d) shows the ultrasonic attenuation coefficient of $d$-wave superconductors with RPA, with results showing a similar behavior quadrupole relaxation rate. Our results indicate the ultrasonic attenuation peak just below $T_{c}$ in cuprates is the coherence peak in charge channel \cite{Xu1988, Sun1989}.

As discussed above, the coherence factor is closely related to the appearance of the coherent peak, primarily due to the sign of $\Delta_k\Delta_{k+q}$. In cuprates, the opposite signs of order parameter at $q=(\pi,\pi)$ due to $d$-wave pairing symmetry result in a strong coherent peak. A natural question arises: do similar phenomena occur in other superconducting systems? To explore this, we investigate the high-temperature iron-based superconductors, which are widely believed to exhibit $s_{\pm}$-wave pairing with sign-changing superconducting order parameters \cite{iron1,iron2,iron_review}.

\begin{figure}[t]
	\begin{center}
		\fig{3.4in}{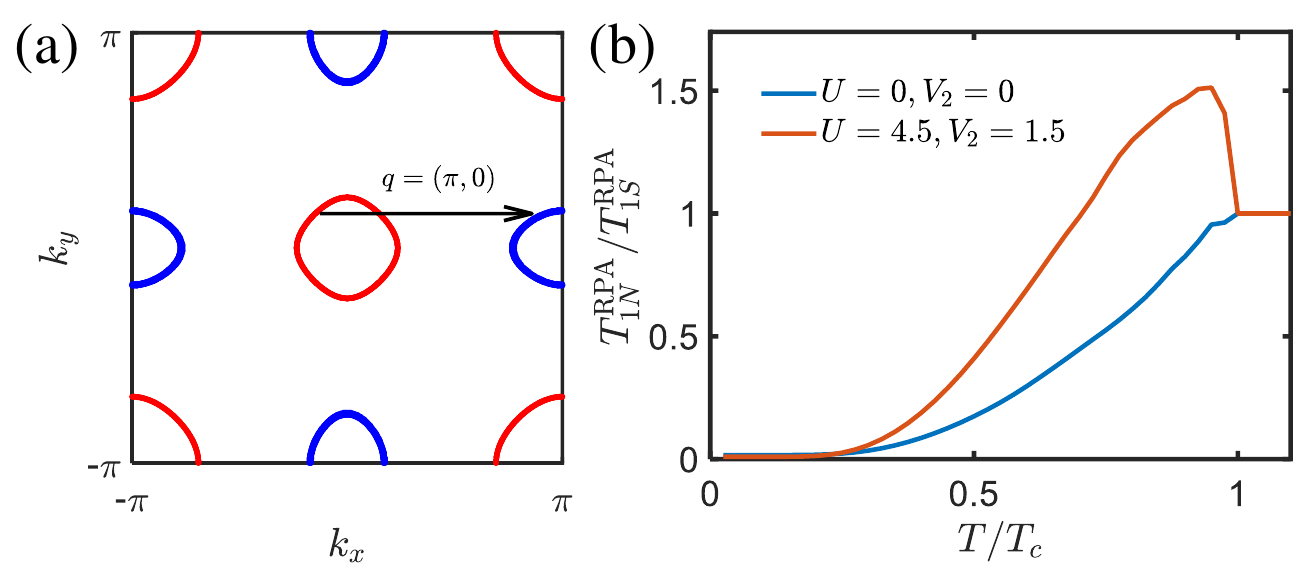}
		\caption{(a) The Fermi surface of the iron-based superconductor two orbital model Eq.\ref{iron} at $\mu=1.4$. The color indicates the sign of the order parameter on each FS, where $q=(\pi,0)$ becomes a dominant momentum.
        (b) The quadrupole relaxation rate of iron-based superconductors calculated based on charge susceptibility for the bare noninteracting case (blue line) and for the RPA renormalized case (orange line). A coherence peak shows up at the orange line below $T_c$.
			\label{fig3}}
	\end{center}
		\vskip-0.5cm
\end{figure}

To simplify our discussion, we use the two-band exchange model proposed in \cite{iron_twoband} to calculate the quadrupole-lattice relaxation rate. Considering two orbitals per site to capture the degeneracy of $d_{xz}$ and $d_{yz}$ orbitals of Fe atom near the Fermi surface, the Hamiltonian of the normal state can be written as
\begin{equation}
H_0=\sum_{\mathbf{k} \sigma} \psi_{\mathbf{k} \sigma}^{\dagger}\left(\begin{array}{cc}
\xi_x(\mathbf{k}) & \epsilon_{x y}(\mathbf{k}) \\
\epsilon_{x y}(\mathbf{k}) & \xi_y(\mathbf{k})
\end{array}\right) \psi_{\mathbf{k} \sigma},
\label{iron}
\end{equation}
where $\psi_{\mathbf{k} \sigma}^{\dagger}=\left(c_{1, \mathbf{k}, \sigma}^{\dagger}, c_{2, \mathbf{k}, \sigma}^{\dagger}\right)$ is the creation operator of the two orbitals. The energy dispersions are
$\xi_x(\mathbf{k})=-2 t_1 \cos k_x-2 t_2 \cos k_y-4 t_3 \cos k_x \cos k_y -\mu$, 
$\xi_y(\mathbf{k})=-2 t_1 \cos k_y-2 t_2 \cos k_x-4 t_3 \cos k_x \cos k_y -\mu$, 
$\epsilon_{x y}(\mathbf{k})=-4 t_4 \sin k_x \sin k_y$,
where the paramters are set as $t_1=-1$, $t_2=1.3$, $t_3=t_4=-0.85$ and $\mu=1.4$ for hole doping. Fig.\ref{fig3}(a) shows the FS of the system where the hole pocket and the electron pocket are connected by $q=(\pi,0), (0,\pi)$ \cite{HongDing_2008}. 
To obtain $s_\pm$-wave pairing, we consider only the next nearest neighbor exchange interaction $J_2 \mathbf{S}_i\mathbf{S}_j$. The mean-field Hamiltonian can then be written as $H_{\mathrm{MF}}=\sum_{\mathbf{k}} \Psi_{\mathbf{k}}^{\dagger} h(\mathbf{k}) \Psi_{\mathbf{k}}$ \cite{iron_pairing},
\begin{equation}
h(\mathbf{k})=\left(\begin{array}{cccc}
\xi_x(\mathbf{k})  & \epsilon_{x y}(\mathbf{k}) & \Delta_1(\mathbf{k})  & 0 \\
\epsilon_{x y}(\mathbf{k}) & \xi_y(\mathbf{k}) & 0 & \Delta_2 (\mathbf{k}) \\
 \Delta_1^*(\mathbf{k})  & 0 & -\xi_x(\mathbf{k})  & -\epsilon_{x y}(\mathbf{k}) \\
 0 & \Delta_2^*(\mathbf{k}) & -\epsilon_{x y}(\mathbf{k}) &  -\xi_y(\mathbf{k})
\end{array}\right),
\end{equation}
where $\Delta_1(\bk)=\Delta_2(\bk)=4\Delta_0\cos{k_x}\cos{k_y}$ is the $s\pm$-wave pairing order parameter obtained by self-consistent calculation. 

The bare charge susceptibility under orbital basis of multi-orbital system can be defined as 
\begin{equation}
\chi_{o_1o_2;o_3o_4}^{0}(q, i \omega)=\int_0^\beta d \tau\left\langle T_\tau \rho_{o_1o_2}(q, \tau) \rho_{o_3o_4}(-q, 0)\right\rangle e^{i \omega \tau} \nonumber
\end{equation}
where $o_{1-4}=1, 2$. Within RPA, we consider the onsite Coulomb repulsion of both intra-orbital repulsion $U$ and inter-orbital repulsion $U'=U$ and nearest neighbor repulsion $V_1$ and next nearest neighbor repulsion $V_2$. 
Then the interaction vertex can be written into $4\times4$ identical matrix form $\Gamma=(U+V(q))I_4$ with $V(q)=2V_1(\cos{q_x}+\cos{q_y}) + 4V_2\cos{q_x}\cos{q_y}$. Then the charge susceptibility with RPA is given by
\begin{equation}
    \chi_{o_1o_2;o_3o_4}^{\mathrm{RPA}}(q, i \omega)=\left[\mathbf{I}-\Gamma \chi_{o_1o_2;o_3o_4}^{0}(q, i \omega)\right]^{-1} \chi_{o_1o_2;o_3o_4}^{0}(q, i \omega) \nonumber
\end{equation}
With the RPA charge susceptibility, we can obtain the quadrupole relaxation rate of the iron-based superconductors shown in Fig.\ref{fig3}(b).
For the bare relaxation rate, the peak is extremely weak owing to the $q\rightarrow 0$ part. When $U=4.5$ and $V_2=1.5$ correlation effect is included, the coherence peaks show up below $T_c$ as in the $d$-wave superconductor. Therefore, the iron-based superconductor could also host a coherence peak in the charge channel.

On the other hand, experimentally achieving measurable quadrupole relaxation always requires a finite EFG and asymmetry parameter $\eta$ \cite{NQR_PhysRevX.14.041072,nqr_mitchell}.
However, $\eta$ is quite small for most iron-based superconductors. On the other hand, ultrasonic attenuation is a promising way for this coherence peak.
It is evidence that a peak just below $T_{c}$ in ultrasonic attenuation \cite{Kurihara2017, Saint2012}. We argue that the peak observed is the same coherence peak in quadrupole relaxation.



In conclusion, we present a comprehensive study of coherence peaks in unconventional superconductor charge channels. Upon entering the superconducting phase at $T_c$, an unconventional coherence peak is identified. This peak is intrinsically linked to the logarithmic divergence in the density of states and the momentum-dependent coherence factor $g(k,q)$ characteristic of $d$-wave superconductors. In contrast, for conventional $s$-wave superconductors, the coherence factor $g(k,q)$ leads to a rapid drop feature instead of a peak.
Nuclear quadrupole relaxation rate offers a direct method to observe this coherence peak, as demonstrated in YBa$_2$Cu$_4$O$_8$
\cite{NQR_PhysRevX.14.041072}. Additionally, ultrasonic attenuation can serve as another effective probe for detecting this feature. Furthermore, we show that a similar coherence peak emerges in iron-based superconductors with $s_{\pm}$ pairing symmetry.
We also want to add a reminder here. Although the coherence peak is observed from the charge fluctuation channel, this charge fluctuation has nothing to do with the pairing mechanism. 
We hope our findings provide valuable insights and open new avenues for understanding coherence peaks in unconventional superconductors.

\textit{Acknowledgement:}
We acknowledge the support by the National Key Research and Development Program of China (Grants No. 2022YFA1403903, No. 2022YFA1602800), the National Natural Science Foundation of China (Grant NSFC-12494590, No. NSFC-12134018, No. NSFC-12174428, and No. NSFC-12274279), the New Cornerstone Investigator Program, and the Chinese Academy of Sciences Project for Young Scientists in Basic Research (2022YSBR-048). 

\bibliography{reference}

\end{document}